\documentclass[11pt]{article}

\usepackage{epsf}
\usepackage{epsfig}
\usepackage{cite}
\usepackage{amsmath}
\usepackage{graphics}

\begin{document}

\setlength{\textheight}{21.5cm}
\setlength{\oddsidemargin}{1.cm}
\setlength{\evensidemargin}{1.cm}
\setlength{\topmargin}{0.cm}
\setlength{\footskip}{1cm}
\setlength{\arraycolsep}{2pt}

\renewcommand{\thefootnote}{\#\arabic{footnote}}
\setcounter{footnote}{0}

\newcommand{\gtrsim}{ \mathop{}_{\textstyle \sim}^{\textstyle >} }
\newcommand{\lesssim}{ \mathop{}_{\textstyle \sim}^{\textstyle <} }
\newcommand{\rem}[1]{{\bf #1}}
\renewcommand{\thefootnote}{\fnsymbol{footnote}}
\setcounter{footnote}{0}
\def\thefootnote{\fnsymbol{footnote}}

\begin{center}

{\Large \bf Atmospheric Neutrino Octant \\from Flavour Symmetry}

\vskip .45in

{\bf  $^1$Paul H. Frampton\footnote{paul.h.frampton@gmail.com}
 $^1$Claudio Corian\`o\footnote{claudio.coriano@le.infn.it}  and $^2$Pietro Santorelli
 \footnote{pietro.santorelli@na.infn.it}}

\vskip .3in

{$^1$Dipartimento di Matematica e Fisica ``Ennio De Giorgi", \\
Universit\`a del Salento and INFN Lecce, Via Arnesano 73100 Lecce, Italy\\}
\vspace{0.4cm}
{$^2$ Dipartimento di Fisica "Ettore Pancini", Universit\`a degli studi di Napoli
"Federico II", Complesso Univ. Monte S. Angelo, I-80126 Napoli, Italy\\
and INFN - Sezione di Napoli, Complesso Univ. Monte S. Angelo, I-80126 Napoli, Italy}

\end{center}

\vskip .4in 
\begin{abstract}
\noindent
The binary tetrahedral group ($T^{'}$) has provided the most successful
flavour symmetry in understanding simultaneously the three mixing angles 
both for quarks in the CKM matrix and for neutrinos in the PMNS matrix. One 
prediction, invariant under
leptonic CP violation, relates the atmospheric
and reactor neutrino mixings $\theta_{23}$ and $\theta_{13}$ respectively.
We study sedulously this relation using the latest neutrino data.
It is natural to focus on the frustrating experimental octant ambiguity
of $\theta_{23}$. We conclude that the flavour
symmetry requires that $\theta_{23}$ is in the second
octant $\theta > 45^0$, not in the first one $\theta_{23}<45^0$, and eagerly
await experimental confirmation of this prediction.

\end{abstract}

\renewcommand{\thepage}{\arabic{page}}
\setcounter{page}{1}
\renewcommand{\thefootnote}{\#\arabic{footnote}}

\newpage

\section{Introduction}

\noindent
In this paper we shall discuss an application of the finite non-abelian
group $T^{'}$, of order $g=24$, to the problem of neutrino mixing between
the three neutrinos $\nu_e, \nu_{\mu}, \nu_{\tau}$, in particular to the
important unanswered question of whether the atmospheric mixing angle
$\theta_{23}$ is in the first octant $0 < \theta_{23} < 45^0$ or in the
second octant $45^0 < \theta_{23} < 90^0$.

\noindent
The group $T^{'}$ is often introduced as the double-cover of the
group $T$ which is simpler to visualise as the 
symmetry of a tetrahedron which is easily seen to coincide
with the even elements of the symmetric group $S_4$. The relationship
of $T^{'}$ to $T$ as a double-cover is strictly analogous to that of $SU(2)$ to
$SO(3)$. There is a subtlety, however, in that the relationship is
not one of a subgroup because of the fact\cite{FKR} that
$SO(3)\not\subset SU(2)$ and $T \not\subset T^{"}$.

\noindent
This subtlety is a global property of these groups which does not
interfere with local properties such as the table of Kronecker
products as can be seen in multiplication tables for the
irreducible representations of $T(1,1^{'},1^{"}; 3)$ and
$T^{'}(1,1^{'},1^{"}; 2,2^{'},2^{"}; 3)$, respectively.

\begin{table}[h]
\caption{Multiplication table for irreducible representations of $T$}
\begin{center}
\begin{tabular}{||c||c|c|c|c||}
\hline
\hline
   & $1$ & $1^{'}$ & $1^{"}$ & $3$   \\
\hline
\hline
$1$  & $1$& $1^{'}$ & $1^{"}$ & 3 \\
\hline
$1^{'}$ & $1^{'}$ & $1^{"}$ & $1$  & $3$ \\
\hline
$1^{"}$ & $1^{"}$ & $1$ & $1^{'}$ & $3$ \\
\hline
$3$ & $3$ & $3$ & $3$ & $1+1^{'}+1^{"}+3+3$ \\
\hline
\hline
\end{tabular}
\end{center}
\label{Ttable}
\end{table}

\begin{table}[h]
\caption{Multiplication table for the irreducible representations of $T^{'}$}
\begin{center}
\begin{tabular}{||c||c|c|c|c|c|c|c||}
\hline
\hline
  & $1$ & $1^{'}$ & $1^{"}$ &  $2$&$2^{'}$ &$2^{"}$ & $3$  \\
 \hline
 \hline
  $1$ &$1$ & $1^{'}$ & $1^{"}$ & $2$ & $2^{'}$& $2^{"}$ & $3$  \\
\hline
 $1^{'}$ & $1^{'}$  & $1^{"}$ & $1$ & $2^{'}$ & $2^{"}$ & $2$ & $3$    \\
\hline
$1^{"}$ & $1^{"}$  & $1$ & $1^{'}$ & $2^{"}$ & $2$ & $2^{'}$ &  $3$ \\
\hline
$2$ & $2$ & $2^{'}$ & $2^{"}$ & $1+3$ & $1^{'}+3$ & $1^{"}+3$ & $2+2^{'}+2^{"}$ \\
\hline
 $2^{'}$ & $2^{'}$ & $2^{"}$ & $2$ & $1^{'}+3$ & $1^{"}+3$ & $1+3$ & $2+2^{'}+2^{"}$  \\
\hline
 $2^{"}$ & $2^{"}$ & $2$ & $2^{'}$ & $1^{"}+3$ & $ 1+3$ & $1^{'}+3$ &  $2+2^{'}+2^{"}$   \\
\hline
 $3$ & $3$ & $3$ & $3$ & $2+2^{'}+2^{"}$ &  $2+2^{'}+2^{"}$ & $2+2^{'}+2^{"}$ & $1+1^{'}+1^{"}+3+3$   \\
\hline
\hline
\end{tabular}
\end{center}
\label{Tprimetable}
\end{table}

\noindent
We note that omitting the rows and columns corresponding to doublets
from Table(\ref{Tprimetable}) leads precisely to Table(\ref{Ttable}) despite
the fact that globally $T^{'} \not\supset T$.

\noindent
Curiously, it was $T^{'}$ as in Table 2 which was the first of these two attractive flavour
symmetries to be used\cite{FK1995} in particle theory, at a time when
neutrino masses, and hence the PMNS mixing matrix, were not of
interest. This makes sense because the doublets of $T^{'}$ are necessary
to achieve the correct CKM matrix.

\section{T and T-prime}

\noindent
When neutrino masses and oscillations were established, starting
in 1998, the PMNS matrix became the centre of attention. For 
the neutrinos, only triplets and singlets are necessary and so
the smaller group $T$ as in Table 1 achieved great success
\footnote{Note that $T$ was often called $A(4)$.} especially
in the hands of Ernest Ma and others\cite{Ma1,Ma2,Ma3}.

\noindent
To accommodate quarks and the CKM matrix it was most
useful to extend $T$ of Table 1 to $T^{'}$ of Table 2  because
the successes of $T$ could be retained and, by using the
$T^{'}$ doublets and a $(2+1)$ family structure, one could
the achieve excellent fits to the mixing angles
of the CKM matrix\cite{Carr,FKM,EFM,EF} by the
Chapel Hill group.

\noindent
One of the outstanding questions in neutrino physics is
the correct octant of the PMNS angle $\theta_{23}$
which remains an ambiguity even from the latest experimental data.
This can be, and has already been\cite{Frampton},
addressed using the $T^{'}$ flavour symmetry.

\noindent
Two other outstanding questions in neutrino
physics
are the hierarchy issue, and the CP violation phase.
As far as we can see, both of these pressing issues
cannot be resolved even by $T^{'}$ flavour symmetry
without more input from experiment. For the latter CP
issue, an important follow-up question will be whether
the phase is related to the phase necessary for generating
matter-antimatter asymmetry in leptogenesis,
as possible only in the FGY model\cite{FGY}.

\noindent
The goal of the present article is to revisit the
octant ambiguity discussed in 2017\cite{Frampton}
in the light of much new experimental data which
we take from the November 2022 entry of \cite{Experiment},
and in the light of all the interest in this ambiguity
\cite{Branco2017,Singh,Das,Huang,Smirnov,Haba,Chakrabarty,Devi,Chaterjee}.

\noindent
There remains the question therefore of an octant prediction from $T^{'}$
using the much more precise recent experimental data on neutrinos, as 
we could use 
it as a black-white litmus test. If this prediction is correct, it
with bolster our confidence even further in this flavour symmetry.
If it were incorrect, it would cast doubt on all the successes
with the six PMNS and CKM mixing angles in \cite{Carr,FKM,EFM,EF}.
We recall that of the many free parameters in the Standard Model,
these 6 are successfully predicted by $T^{'}$ flavour symmetry while
 the other 22 free parameters are
not yet predicted by anything. Having said that, this impressive flavour 
symmetry has not so far made any progress with fermion 
masses, although not from lack of trying.

\section{Comparison to latest neutrino data}

\noindent
The most robust $T^{'}$ prediction, invariant under leptonic CP
violation, is this one, relating $\theta_{13}$ with $\theta_{23}$:

\begin{equation}
\theta_{13} = \sqrt{2} \left| \frac{\pi}{4} - \theta_{23} \right|
\label{ThePrediction}
\end{equation}

\noindent
We can evaluate the LHS and RHS of Eq. (\ref{ThePrediction})
using the November 2022 data from \cite{Experiment}, as follows.
We use the $3\sigma$ data.

\noindent
The LHS is 0.131 for the first octant. It is 0.148 for the second octant.

\noindent
The RHS of Eq. (\ref{ThePrediction})  is in the range from 0.143 to 0.156.

\noindent
The unique $T^{'}$ prediction is therefore that $\theta_{23}$ is in the second octant.
This agrees with the 2017 result \cite{Frampton} which used less precise data.

\noindent
We should point out that if we restrict attention to only the leptonic
sector, Eq. (\ref{ThePrediction}) can actually be derived using $T$
flavour symmetry. This is an academic remark, because successfully
to include quarks, and the CKM matrix, we must use $T^{'}$ flavour
symmetry.

\section{Discussion}

\noindent
Therefore  we stick our necks out about the $T^{'}$ flavour
symmetry, buoyed by its success \cite{Carr,FKM,EFM,EF}
with all six of the PMNS and CKM mixing angles. We predict
confidently that $\theta_{23}$ is in the second octant. 
If this turns out experimentally to be the correct resolution
of the octant ambiguity, we shall gain even more confidence
in $T^{'}$. If this prediction is refuted experimentally, we
shall admit that $T^{'}$ is fatally flawed, and never discuss
it again.

\noindent
As for the hierarchy problem which depends on the
sign of $\triangle m^2_{23}$, all we can say is that most
model building leads to a normal hierarchy which
leads us to suspect that normal is more likely than
the inverted hierarchy. But this is admittedly 
prejudice and only experiment can resolve this
ambiguity by establishing the sign  for  $\triangle m^2_{23}$.

\noindent
Regarding the CP-violating phase $\delta_{CP}$ in the PMNS matrix,
the present prejudice is $\delta_{CP} \simeq - 90^0$ although
we do not really know this.

\noindent
Perhaps the deepest question in neutrino physics is whether $\delta_{CP}$
is related to $\delta_{LG}$, the CP-violating phase occurring within
leptogenesis in the decay of right-handed neutrinos. In a general model,
this is not the case. The phase $\delta_{LG}$ is all important for establishing
the correct mechanism involved in matter-antimatter asymmetry.
In one very special case\cite{FGY} there is  a direct
connection between $\delta_{LG}$ and $\delta_{CP}$. If this is
the case in Nature, the long-baseline neutrino experiments
which will measure $\delta_{CP}$ are simultaneously shedding
light on one of the greatest mysteries in the early universe,
matter-antimatter asymmetry.

\noindent
Establishing such a relationship requires knowledge of the
right-handed neutrinos which are possibly super-heavy and
this renders establishing such a linkage extremely challenging
at least in the foreseeable future\footnote{A similar remark
is equally valid for the see-saw model \cite{Minkowski}
of neutrino masses.}.


\begin{thebibliography}{99}
\bibitem{FKR}
P.H. Frampton, T.W. Kephart and R.M Rohm,\\
Phys. Lett. {\bf B679,} 478 (2009).\\
{\tt arXiv:0904.0420[hep-ph]}.
\bibitem{FK1995}
P.H. Frampton and T.W. Kephart,\\
Int. J. Mod. Phys.  {\bf A10,} 4689 (1995).\\
{\tt arXiv:hep-ph/9409330}.
\bibitem{Ma1}
E. Ma and G. Rajasekaran,\\
Phys. Rev. {\bf D64,} 113012 (2001).\\
{\tt arXiv:hep-ph/0106291[hep-ph]}
\bibitem{Ma2}
K.S. Babu, E. Ma and J.F.W. Valle,\\
Phys. Lett. {\bf B522,} 207 (2003).\\
{\tt arXiv:hep-ph/0206292[hep-ph]}.
\bibitem{Ma3}
E. Ma,\\
Phys. Rev {\bf D70,} 031901 (2004),\\
{\tt arXiv:hep-ph/0404199[hep-ph]}.
\bibitem{Carr}
P.~D.~Carr and P.~H.~Frampton,\\
{\tt arXiv:hep-ph/0701034.}
\bibitem{FKM}
P.H. Frampton, T.W. Kephart and S. Matsuzaki,\\
Phys. Rev. {\bf D78,} 073004 (2008). \\
{\tt arXiv:0807.4713[hep-ph]}.
\bibitem{EFM}
D.A. Eby, P.H. Frampton and S. Matsuzaki,\\
Phys. Lett. {\bf B671,} 386 (2009).\\
 {\tt arXiv:0810.4899[hep-ph]}.
\bibitem{EF}
D.A. Eby and P.H. Frampton, \\
Phys. Rev. {\bf D86,} 117304 (2012).\\
{\tt arXiv:1112.2675[hep-ph]}.
\bibitem{Frampton}
P.H. Frampton,\\
Mod. Phys. Lett. {\bf A32,} 1750101 (2017).\\
{\tt arXiv:1707.01383[hep-ph]}.
\bibitem{FGY}
P.H. Frampton, S.L.Glashow and T. Yanagida,\\
Phys. Lett. {\bf B548,} 119 (2002).\\
{\tt arXiv:hep-ph/0208157}.
\bibitem{Experiment}
http://www.nu-fit.org
\bibitem{Branco2017}
G.C. Branco, M.N. Rebelo, J.I. Silva-Marcos.\\
JHEP {\bf 11:}001 (2017) \\
{\tt arXiv:1705.07758 [hep-ph]}.
\bibitem{Singh}
M. Singh,\\
{\tt arXiv:1706.06473[hep-ph]}.
\bibitem{Das}
C.R. Das, J. Maalampi, S. Vihonen and J. Pulido,\\
Phys. Rev. {\bf D97,} 035032 (2018).\\
{\tt arXiv:1708.05182[hep-ph]}.
\bibitem{Huang}
G.-Y. Huang, Z.-Z. Xing and J.-Y. Zhu,\\
{\tt arXiv:1806.06640[hep-ph]}.
\bibitem{Smirnov}
M.V. Smirnov, Z.J. Hu, S.J. Li and J.J. Ling,\\
Chinese Phys. {\bf C43,} 033001 (2019).\\
{\tt arXiv:1808.03795[hep-ph]}.
\bibitem{Haba}
N. Haba, Y. Mimura and T. Yamada,\\
Phys. Rev. {\bf D101,} 075034 (2020).\\
{\tt arXiv:1812.10940[hep-ph]}.
\bibitem{Chakrabarty}
K. Chakrabarty, S. Goswami, C. Gupta and T. Thakore,\\
JHEP {\bf 05:}137 (2019).\\
{\tt arXiv:1902.02963[hep-ph]}.
\bibitem{Devi}
M.R. Devi and K.Bora,\\
Mod.Phys. Lett. {\bf A37,} 2250073  (2022).\\
{\tt arXiv:2112.13004[hep-ph]}.
\bibitem{Chaterjee}
A. Chaterjee, S. Goswami and S. Pan,\\
{\tt arXiv:2212.02949[hep-ph]}.
\bibitem{Minkowski}
P. Minkowski,\\
Phys. Lett. {\bf B67,} 421 (1977).


\end{thebibliography}
\end{document}